\documentclass[11pt,a4paper]{article}
\pdfoutput=1
\usepackage{graphicx}
\usepackage{pict2e} 
\usepackage{diagbox} 
\usepackage{setspace}
\usepackage{amsfonts}
\usepackage{jheppub}
\usepackage{amsmath}
\usepackage{amssymb}
\usepackage{color}
\usepackage{float}
\usepackage[normalem]{ulem}
\newcommand{\stkout}[1]{\ifmmode\text{\sout{\ensuremath{#1}}}\else\sout{#1}\fi}
\allowdisplaybreaks

\title{Energy functionals from Conformal Gravity}

\author[a,b]{Giorgos Anastasiou,}
\author[a,b]{Ignacio J. Araya,}
\author[c]{Rodrigo Olea}

\affiliation[a]{Instituto de Ciencias Exactas y Naturales (ICEN), Universidad Arturo Prat, Avenida Arturo Prat Chac\'on 2120, 1110939, Iquique, Chile}
\affiliation[b]{Facultad de Ciencias, Universidad Arturo Prat, Avenida Arturo Prat Chac\'on 2120, 1110939, Iquique, Chile}
\affiliation[c]{Departamento de Ciencias F\'isicas, Universidad Andres Bello,\\
Sazi\'e 2212, Piso 7, Santiago, Chile.}

\emailAdd{ganastasiou@unap.cl, ignaraya@unap.cl, rodrigo.olea@unab.cl}

\abstract{We provide a new derivation of the  Hawking mass and Willmore energy functionals for asymptotically AdS spacetimes, by embedding Einstein-AdS gravity in Conformal Gravity. By construction, the evaluation of the four-dimensional Conformal Gravity action in a manifold with a conical defect produces a codimension-2 conformal invariant functional $L_{\Sigma}$. The energy functionals are then particular cases of $L_{\Sigma}$ for Einstein-AdS and pure AdS ambient spaces, respectively. The bulk action is finite for AdS asymptotics and both Hawking mass and Willmore energy are finite as well. The result suggests a generic relation between conformal invariance and renormalization, where the codimension-2 properties are inherited from the bulk gravity action.}

\begin{document}

\maketitle

\section{Introduction}

Energy functionals play an important role in differential geometry, and specially in the study of embedded extremal surfaces. For example, in biology, the minimization of such functionals associated to mechanical properties of cell membranes, can be used to explain the shape of biological structures under different types of stress \cite{Lomholt_2006,https://doi.org/10.48550/arxiv.1709.04399}. In mathematics, the study of minimal surfaces and of extrema of isoperimetric inequalities or volume maximization, play an important historical role. In this sense, minimal area surfaces appear ubiquitously when predicting the shapes of membranes with surface tension \cite{isenberg1978science,reilly1982mean}. In point of fact, it has been known since antiquity that the sphere and its lower-dimensional analogues, are the forms that maximize volume for a fixed surface area.

Besides the interest of energy functionals and extremal surfaces \textit{per se}, in the context of quantum field theories and many-body systems, Entanglement Entropy (EE) has been regarded as a valuable computational resource \cite{2006PhDT........59E,Horodecki:2009zz,Amico:2007ag}. It has been also considered as an order parameter indicative of phase transitions in quantum systems, of the presence of short or long range correlations, of topological order, etc.\cite{Kitaev:2005dm,Klebanov:2007ws,Levin:2006zz}. It was very interesting then, when in the framework of the anti-de Sitter/Conformal Field Theory (AdS/CFT) correspondence \cite{Maldacena:1997re,Gubser:1998bc,Witten:1998qj}, a relation between the EE of the spatial region of a CFT and the area of a minimal codimension-2 surface in the dual bulk spacetime was found. Indeed, this formula is reminiscent of the Bekenstein-Hawking entropy formula for black holes in Einstein gravity \cite{Ryu:2006ef}, which points to a connection between information and geometry \cite{Maldacena:2013xja,Stanford:2014jda,Balasubramanian:2013lsa}. 

In the saddle-point approximation of the duality, one considers that the on-shell action of the bulk gravity theory defines the generating functional of connected correlators of the corresponding dual CFT \cite{Witten:1998qj,Gubser:1998bc}. In this picture, from the near-boundary expansion of the bulk fields, one identifies both the external source and its canonical conjugate, i.e. the holographic response function of the respective CFT operator \cite{deHaro:2000vlm}. Then, the holographic EE (HEE) measures the amount of entanglement between a spatial region in the CFT --the entangling region-- and its exterior, and can be obtained holographically as a codimension-2 functional which depends on the bulk gravity action and is therefore a geometric object.

In particular, the HEE for Einstein-AdS gravity is well-known to be given by the area of the Ryu-Takayanagi (RT) minimal surface \cite{Ryu:2006ef}, which is anchored at the conformal boundary. This area is divergent due to the pole in the AdS metric. In order to isolate the universal part, standard holographic techniques in codimension-2 have been considered \cite{Taylor:2016aoi}, such that renormalized area is inherited from renormalized bulk gravity action.

In this work we present a prescription to derive a codimension-2 local conformal invariant, named $L_{\Sigma}$, directly from Conformal Gravity (CG) evaluated on the replica orbifold. This object can be applied to obtain the renormalized HEE for Einstein-AdS spacetimes, and also for deriving energy functionals such as Willmore energy \cite{willmore1996riemannian,Toda2017Willmore} and reduced Hawking mass \cite{Fischetti:2016fbh}.

Obtaining the energy functionals from $L_{\Sigma}$ is an example of Conformal Renormalization, where Einstein-AdS gravity is consistently embedded into CG, in order to provide bulk renormalization, as discussed in the four and six-dimensional cases in Ref.\cite{Anastasiou:2020mik}. This hints at a connection between conformal invariance and renormalization of bulk/codimension-2 functionals.

This paper is organized as follows. In section \ref{Section 2}, we consider the renormalized area formula of Ref.\cite{Alexakis:2010zz} and we study its properties under conformal transformation in order to emphasize that it is not a conformal invariant. Then, we restore conformal invariance by constructing $L_{\Sigma}$, that reduces to the renormalized area when evaluated on boundary-anchored minimal surfaces on Einstein-AdS ambient spacetimes. We also relate $L_{\Sigma}$ to the Willmore energy. In section \ref{CGconical}, we obtain $L_{\Sigma}$ directly from the evaluation of the CG action on a conically-singular manifold. In section \ref{ReducedHawking}, we use $L_{\Sigma}$ to derive the reduced Hawking mass \cite{Fischetti:2016fbh}, by evaluating the former on Einstein-AdS ambient spacetimes. Finally, in section \ref{Conclusions}, we conclude with a summary of our results.

\section{Obtaining $L_{\Sigma}$: the hard way}\label{Section 2}

The RT formula \cite{Ryu:2006ef} for the computation of the holographic entanglement entropy (HEE) opened a new window in the analysis of energy functionals and its properties. Most importantly, it were Lewkowycz and Maldacena \cite{Lewkowycz:2013nqa} who proved the conjectured RT formula and provided a systematic way to derive HEE functionals. There, the holographic application of the replica trick induces an orbifold structure in the bulk in the presence of conical singularities due to the replica symmetry. The evaluation of the Einstein-Hilbert action in the orbifold make manifest the area functional located at the cosmic brane where the conical singularities are located. Indeed, for the conical expansion of the Ricci scalar we get

\begin{equation}
\int\limits_{M^{\left(\alpha\right)}} d^{4}x \sqrt{g} R^{\left(\alpha\right)}=\int\limits_{M} d^{4}x\sqrt{g}R + 4 \pi \left(1-\alpha\right) \mathcal{A} \left[\Sigma\right]\,, \label{RicciFPS}
\end{equation}
where $\mathcal{A} \left[\Sigma\right]=\int\limits_{\Sigma} d^{2}y \sqrt{\gamma} $ is the area functional.

It becomes clear from the previous construction that the area can be naturally extracted out of the Ricci scalar in the presence of cones \cite{Fursaev:2013fta}. At this point, one should wonder whether the renormalized codimension-2 area arises out of the renormalized bulk EH action. Indeed, earlier works \cite{Taylor:2016aoi} indicate that this is the case, applying standard holographic renormalization techniques.

\subsection{Topological terms in four-dimensional AdS gravity}

Of particular relevance for the current discussion is the case of four dimensions, where the terms required for the background independent renormalization of the action, i.e. the holographic renormalization counterterms, have been extensively studied \cite{deHaro:2000vlm,Henningson:1998ey}. In this case, the introduction of the counterterms is equivalent to the addition of a boundary term with a fixed coupling constant and explicit dependence on both the intrinsic and extrinsic curvature, dubbed second Chern form $B_3$. The latter works as a resummation of the standard counterterms \cite{Miskovic:2009bm,Anastasiou:2020zwc} and the corresponding renormalized Einstein-AdS action reads 

\begin{equation}
I_{\text{ren}}= \frac{1}{16 \pi G_{N}} \int\limits_{M} d^{4}x  \sqrt{g} \left(R-2 \Lambda \right) + \frac{\ell^2}{64\pi G_{N}} \int\limits_{\partial M} d^{3}x\,B_{3}  \,,
\label{Iren}
\end{equation}
with the cosmological constant defined in terms of the AdS radius as $\Lambda=-\frac{3}{\ell^2}$, and the surface terms given by

\begin{equation}
B_{3}= 4  \sqrt{h}\, \delta^{a_{1} a_{2} a_{3}}_{b_{1} b_{2} b_{3}} K^{b_{1}}_{a_{1}} \left(\frac{1}{2} \frak{R}^{b_{2} b_{3}}_{a_{2} a_{3}} \left(h\right) - \frac{1}{3} K^{b_{2}}_{a_{2}}K^{b_{3}}_{a_{3}}\right) \,.
\label{chernform}
\end{equation}
Here $g_{\mu \nu}$ and $h_{ab}$ are the bulk and the boundary metric, respectively, where the Greek indices denote bulk coordinates and the Latin indices $(a,b)$ correspond to boundary coordinates (see Appendix \ref{AppendixA} for conventions). In this respect, adding up the Chern form to the bulk Lagrangian provides a more geometric approach to the problem of renormalization of AdS gravity.  Indeed, this term arises as the boundary correction to the Euler characteristic in the Gauss-Bonnet theorem for non-compact manifolds, i.e.,

\begin{equation}
\int\limits_{M} d^{4}x\,\mathcal{E}_{4} = 32 \pi^2 \chi \left[M\right] + \int\limits_{\partial M} d^{3}x\,B_{3} \,,
\label{Eulertheorem}
\end{equation}
where

\begin{equation}
\mathcal{E}_{4} =  \sqrt{g} \left(Rie^2 - 4 Ric^2+ 4 R^{2}\right) \,,
\label{GaussBonnet}
\end{equation}
and $\chi\left[M\right]$ is the Euler characteristic of the manifold $M$.
An immediate consequence of the above relation is the fact that the use of extrinsic counterterms is equivalent, up to the Euler characteristic, to the addition of a topological invariant of the Euler class. This implies

\begin{equation}
I_{\text{ren}} = \frac{1}{16 \pi G_{N}} \int\limits_{M} d^{4}x  \sqrt{g} \left(R-2 \Lambda \right) +  \frac{\ell^2}{64 \pi G_{N}} \int\limits_{M} d^{4}x\,\mathcal{E}_{4} -\frac{\pi \ell^2}{2 G_{N}} \chi \left[M\right] \,.
\label{IrenGB}
\end{equation}
The main advantage of this prescription is that one can interchange a boundary for a bulk term, since they are locally equivalent. It also unveils topological features of the corresponding manifold, captured by the topological number $\chi \left[M\right]$.

\subsection{Renormalized area from topological terms and conical defects}

The generalized form of gravitational entropy considers the evaluation of the corresponding bulk Lagrangian in a squashed conically singular manifold \cite{Lewkowycz:2013nqa}.
Along this line, Fursaev, Patrushev and Solodukhin (FPS) in Ref.\cite{Fursaev:2013fta} analyzed the behavior of quadratic terms in the curvature in the vicinity of a squashed cone with an angular deficit $2\pi(1-\alpha)$. This analysis implies that the square of the Riemann tensor is decomposed as

\begin{align}
&\int\limits_{M^{\left(\alpha\right)}} d^{4}x \sqrt{g}
\left(Rie^{\left(\alpha\right)}\right)^{2} =\int\limits_{M} d^{4}x\sqrt{g}\, Rie^{2} \nonumber\\
& + 8\pi \left(1-\alpha\right) \int\limits_{\Sigma} d^{2}y \sqrt{\gamma} \left(R_{ABAB} - \mathcal{K}^{\left(A\right)}_{ij} \mathcal{K}^{ij}_{\left(A\right)}\right) + \mathcal{O}\left(\left(1-\alpha\right)^2\right) \,, \label{Riemsquared}
\end{align}
while the Ricci tensor squared reads

\begin{align}
&\int\limits_{M^{\left(\alpha\right)}} d^{4}x \sqrt{g} \left(Ric^{\left(\alpha\right)}\right)^{2}= \int\limits_{M} d^{4}x \sqrt{g}\,  Ric^{2} \nonumber \\
& + 4\pi \left(1-\alpha\right) \int\limits_{\Sigma} d^{2}y \sqrt{\gamma} \left(R_{AA} -\frac{1}{2} \left(\mathcal{K}^{\left(A\right) }\right)^{2}\right) +\mathcal{O}\left(\left(1-\alpha\right)^2\right) \,. \label{Ricsquared}
\end{align}
In turn, the square of the Ricci scalar  splits as follows

\begin{align}
& \int\limits_{M^{\left(\alpha\right)}} d^{4}x \sqrt{g} \left(R^{\left(\alpha\right)}\right)^{2} = \int\limits_{M} d^{4}x \sqrt{g}\, R^{2} + 8\pi \left(1-\alpha\right) \int\limits_{\Sigma}d^{2}y \sqrt{\gamma}\,R + \mathcal{O}\left(\left(1-\alpha\right)^2\right) \,, \label{RicciScalaquared}
\end{align}
while the Euler characteristic is expanded as

\begin{equation}
\chi \left[M^{\left(\alpha\right)}\right] = \chi \left[M\right]+ \left(1-\alpha\right) \chi\left[\Sigma\right]\,.\label{EulercharFPS}
\end{equation}
Here $M^{\left(\alpha\right)}$ represents the four-dimensional orbifold, and $\Sigma$ is the codimension-2 manifold located at the tip of the conical singularity which is described by the embedding function $x^{\mu}=x^{\mu} \left(y^{i}\right)$ and the induced metric $\gamma_{ij}$, where the Latin indices $(i,j)$ denote codimension-2 coordinates. The labels $\left(A,B\right)$ denote the orthogonal directions to $\Sigma$ and $\mathcal{K}^{\left(A\right)}$ is its extrinsic curvature along the normal direction $n^{\left(A\right)}$. The bulk curvature terms at the r.h.s. of the above relations correspond to the regular part of them.

Taking into account Eqs.\eqref{Riemsquared}-\eqref{RicciScalaquared}, the Gauss-Bonnet term in the presence of squashed cones, adopts the form

\begin{equation}
\int\limits_{M^{\left(\alpha\right)}} d^{4}x\, \mathcal{E}_{4}^{\left(\alpha\right)} = \int\limits_{M} d^{4}x\, \mathcal{E}_{4} + 8 \pi \left(1-\alpha\right) \int\limits_{\Sigma} d^{2}y\, \mathcal{E}_{2} \,, \label{GBFPS}
\end{equation}
where $\mathcal{E}_{2}=\sqrt{\gamma} \mathcal{R}$ is the corresponding topological term in two dimensions. The fact that extrinsic curvatures are not present in the codimension-2 functional is a remarkable feature of the addition of a topological term in the bulk. It also brings in topological contributions to quantum information theoretic measures, e.g., in the context of HEE.

In sum, when the renormalized Einstein-AdS action in Eq.\eqref{IrenGB} is evaluated on the orbifold $M^{\left(\alpha\right)}$,  we get

\begin{equation}
I_{\text{ren}}^{\left(\alpha\right)} = I_{\text{ren}} +\frac{\left(1-\alpha\right)}{4G_{N}} \left(\mathcal{A} \left[\Sigma\right] +\frac{\ell^2}{2} \int\limits_{\Sigma} d^{2}y\, \mathcal{E}_{2} - 2 \pi \ell^2 \chi \left[\Sigma\right]\right) \,.
\end{equation}
Note that the term that is linear in $\left(1-\alpha\right)$ can be identified with the renormalized area given by Alexakis and Mazzeo in Ref.\cite{Alexakis:2010zz}. In order to show this, we rewrite the conical contribution as

\begin{align}
\mathcal{A}_{ren}\left [\Sigma \right ]&= \int\limits_{\Sigma} d^{2}y \sqrt{\gamma} +\frac{\ell^2}{2} \int\limits_{\Sigma} d^{2}y\, \mathcal{E}_{2} - 2 \pi \ell^2 \chi \left[\Sigma\right] \nonumber \\
&=\frac{\ell^2}{2} \int\limits_{\Sigma}d^{2}y \sqrt{\gamma} \left(\mathcal{R} + \frac{2}{\ell^2}\right) - 2 \pi \ell^2 \chi \left[\Sigma\right] \nonumber \\
&= \frac{\ell ^{2}}{4}\int\limits_{\Sigma }d^{2}y\sqrt{\gamma}\delta _{ij}^{km}\left (\mathcal{R}_{km}^{ij} +\frac{1}{\ell ^{2}}\delta _{km}^{ij}\right ) -2\pi \ell ^{2}\chi \left [\Sigma \right ] \,.
\label{Aren}
\end{align}
Considering the Gauss-Codazzi relation the latter can be cast as

\begin{align}
\mathcal{A}_{ren}\left [\Sigma \right ]
&= \frac{\ell ^{2}}{4}\int\limits _{\Sigma }d^{2}y\sqrt{\gamma}\delta _{ij}^{ms}\left (R_{ms}^{ij} +2\mathcal{K}_{ms}^{\left (A\right )}\mathcal{K}_{\left (A\right )}^{ij} +\frac{1}{\ell ^{2}}\delta _{ms}^{ij}\right ) -2\pi \ell ^{2}\chi \left [\Sigma \right ] \nonumber\\
&= \frac{\ell ^{2}}{4}\int \limits_{\Sigma }d^{2}y\sqrt{\gamma}\delta _{ij}^{ms}\left (W_{\left(\text{E}\right) ms}^{ij} +2\mathcal{K}_{ms}^{\left (A\right )}\mathcal{K}_{\left (A\right )}^{ij}\right )-2\pi \ell ^{2}\chi \left [\Sigma \right] \,,
\end{align}
where $W_{\left(\text{E}\right) \mu \nu}^{\alpha \beta}$  corresponds to the curvature of the AdS group for (pseudo)Riemannian manifolds without torsion. The same object can also be identified with the Weyl tensor for Einstein spacetimes

\begin{equation}
W_{\left(\text{E}\right) \mu \nu}^{\alpha \beta}=R_{\mu \nu}^{\alpha \beta}+\frac{1}{\ell^{2}}\delta_{ \mu \nu}^{\alpha \beta} \,,
\label{WeylEinstein}
\end{equation}
which comes from the generic form of the Weyl
\begin{equation}
W_{\mu \nu}^{\alpha \beta}=R_{\mu \nu}^{\alpha \beta}- \left(S^{\alpha}_{\mu} \delta^{\beta}_{\nu} -S^{\beta}_{\mu} \delta^{\alpha}_{\nu}-S^{\alpha}_{\nu} \delta^{\beta}_{\mu} +S^{\beta}_{\nu} \delta^{\alpha}_{\mu} \right) \,, \label{Weyltensor}
\end{equation}
where the Einstein condition in the Schouten tensor, $S_{\mu \nu}= -\frac{1}{2 \ell^{2}} g_{\mu \nu}$, is considered.
Finally, in order to express the above formula in a more standard form, one may use the traceless part of the extrinsic curvature, defined as 

\begin{equation}
    P_{ij}^{\left (A\right )} =\mathcal{K}_{ij}^{\left (A\right )} -\frac{1}{2}\mathcal{K}^{\left (A\right )}\gamma _{ij}\,.
\end{equation}
In doing so, one obtains

\begin{equation}
\mathcal{A}_{ren}\left [\Sigma \right ]= \frac{\ell ^{2}}{2}\int\limits_{\Sigma }d^{2}y\sqrt{\gamma}\left [W_{\left(\text{E}\right)ij}^{ij} -P_{ij}^{\left (A\right )}P_{\left (A\right )}^{ij} +2\left (H^{\left (A\right )}\right )^{2}\right ]-2\pi \ell ^{2}\chi \left [\Sigma \right]  \,,
\label{Arenweyl}
\end{equation}
where $H^{\left (A\right )}=\frac{1}{2}\mathcal{K}^{\left(A\right)}$ is the mean curvature of $\Sigma$.

This functional matches the corresponding formula for $\mathcal{A}_{ren}$ given in Ref.\cite{Alexakis:2010zz}.
Hence, the renormalized area naturally arises as the conical contribution of the renormalized Einstein-AdS action in the presence of squashed conical singularities. In a compact notation, the renormalized action on the orbifold takes the form

\begin{equation}
I_{\text{ren}}^{\left(\alpha\right)} = I_{\text{ren}} + \frac{\left(1-\alpha\right)}{4 G_{N}}\mathcal{A}_{\text{ren}} \left[\Sigma\right] \,,
\end{equation}
reflecting the fact that $\mathcal{A}_{ren}$ is inherited from bulk renormalization.
Even though $ I_{\text{ren}}$ is finite for any four-dimensional Einstein spacetime which is asymptotically AdS (AAdS), there are certain subtleties in the class of hypersurfaces that can be renormalized by $\mathcal{A}_{\text{ren}}$. As it was pointed out in Ref.\cite{Alexakis:2010zz}, $\mathcal{A}_{\text{ren}}$ successfully renders finite the area of any minimal or non-minimal surface that is anchored orthogonally to the conformal boundary. However, when the intersection is not orthogonal then the corresponding $\mathcal{A}_{\text{ren}}$ has to be corrected \cite{Fischetti:2016fbh}. Minimal surfaces satisfy trivially this relation, what makes this functional adequate for the calculation of the HEE.

\subsection{Renormalized area is not conformal invariant for arbitrary manifolds}

In the mathematical literature, conformal invariance plays a key role in the definition of Renormalized Volume for asymptotically hyperbolic spacetimes \cite{FG,Graham:1999jg,Albin:2005qka}. As a matter of fact, this bulk functional is expressed in terms of conformal invariants in four \cite{Anderson2000L2CA} and six \cite{Chang:2005ska} dimensions. It is expected that codimension-2 descendants of these structures would be conformally invariant, as well. It is also reasonable to think that this property will give rise to energy functionals for surfaces which are a proper measure of the deviation with respect to extremality (e.g., sphericity, as in the soap bubbles) irrespective of their size.

Taking the above argument as motivation, we will study the behavior of the renormalized area \eqref{Arenweyl} under conformal tranformation of the ambient metric $g_{\mu \nu } =e^{2\phi }\hat{g}_{\mu \nu }$.

The completeness relation 

\begin{equation}
g_{\mu \nu } =n^{\left (A\right )}_{\mu }n_{\left (A\right )\nu } +e^{i}_{\mu }e^{j}_{\nu }\gamma _{i j} \,,
\end{equation}
with $n_{\mu }^{\left (A\right )}$ and $e_{i}^{\mu }$ being the corresponding normal and frame vectors, respectively, dictates the transformation of the following objects under Weyl rescaling, i.e.,

\begin{equation}
n_{\mu }^{\left (A\right )} =e^{\phi }\hat{n}_{\mu }^{\left (A\right )} \,,\ \gamma _{ij} =e^{2 \phi }\hat{\gamma }_{ij} \,.
\label{weyltranformnormalgamma}
\end{equation}
Furthermore, the extrinsic curvature transforms as

\begin{equation}
\mathcal{K}_{ij}^{\left (A\right )} = e^{\phi} \left(\hat{\mathcal{K}}_{ij}^{\left (A\right )} +\hat{\gamma} _{ij} \hat{n}^{\left(A\right)} \partial \phi\right) \,,
\label{Kconformtrans}
\end{equation}
where the contracted indices of $\hat{n}^{\left(A\right)} \partial \phi$ have been omitted for simplicity. Then it is straightforward to show that its trace scales as

\begin{equation}
\mathcal{K}^{\left (A\right )} = e^{-\phi} \left(\hat{\mathcal{K}}^{\left (A\right )} +2 \hat{n}^{\left(A\right)} \partial \phi\right) \,,
\label{Ktrconf}
\end{equation}
leading, in turn, to a conformally covariant $P_{ij}^{\left (A\right )}$ of weight 1. Finally, $W_{\left(\text{E}\right)ij}^{ij}$ transforms as

\begin{equation}
W_{\left(\text{E}\right)\mu \nu}^{\lambda\sigma} =e^{-2\phi} \left[\hat{W}_{\left(\text{E}\right)\mu \nu}^{\lambda \sigma}+\frac{1}{\ell^2} \delta_{\mu \nu}^{\lambda \sigma} \left(e^{2\phi}-1\right)-4\delta_{[\mu}^{[\lambda}\hat{T}_{\nu]}^{\sigma]}\right]\,,
\label{Weyleinstein}
\end{equation}
where
\begin{equation}
\hat{T}_{\nu}^{\lambda}=\hat{\nabla}^{\lambda}\hat{\nabla}_{\nu}\phi-\hat{\nabla}_{\nu}\phi\hat{\nabla}^{\lambda}\phi+\frac{1}{2}\delta_{\nu}^{\lambda}\hat{\nabla}_{\mu}\phi\hat{\nabla}^{\mu}\phi \,.
\end{equation}
Thus, the renormalized area functional \eqref{Arenweyl} transforms as

\begin{align}
\mathcal{A}_{\text{ren}}\left [\Sigma \right ]&= \frac{\ell ^{2}}{2}\int\limits_{\Sigma }d^{2}y\sqrt{\hat{\gamma }}\left[\hat{W}_{\left(\text{E}\right)ij}^{ij} -\hat{P}_{ij}^{\left (A\right ) }\hat{P}_{\left (A\right )}^{ij} +2\left (\hat{H}^{\left (A\right )}\right )^{2} +2\hat{\mathcal{K}}_{\left (A\right )}\hat{n}^{\left (A\right )} \partial \phi \right. \nonumber \\
& \left. +2\left (\hat{n}^{\left (A\right )} \partial \phi \right )^{2} +2 \left(\frac{e^{2\phi}-1}{\ell^2}-\hat{T}_{i}^{i}\right)  \right] -2\pi \ell ^{2}\chi \left [\Sigma \right] \,.
\label{Arenconformalgen}
\end{align}
It is evident from this expression that $\mathcal{A}_{\text{ren}}$ is not symmetric under local conformal transformations when a generic surface is embedded in an arbitrary ambient metric.

\subsection{Restoring conformal invariance in codimension 2}\label{Willmoreenergy}

Rendering Eq.\eqref{Arenweyl} conformally invariant for an arbitrary surface is a  non-trivial task. Indeed, this is equivalent to finding the compensating terms which restore conformal invariance of the renormalized area functional, what is technically involved.

However, the problem gets simpler once specific conditions are imposed in  Eq.\eqref{Arenconformalgen}. In particular, when the ambient metric is an Einstein spacetime both in the physical and in the conformal frame, then $W_{\left(\text{E}\right)ij}^{ij}$ turns conformally covariant with a weight factor -2, as can be seen from Eq.\eqref{Weyleinstein}. This constraint makes the last parentheses in Eq.\eqref{Arenconformalgen} to vanish.

Additionally, considering a minimal embedding surface $\Sigma$, corresponds to the vanishing of the trace of the extrinsic curvature, which can equivalently be expressed in the conformal frame as

\begin{equation}
\hat{\mathcal{K}}^{\left (A\right )} = -2 \hat{n}^{\left(A\right)} \partial \phi \,.
\label{minimalitycondition}
\end{equation}
Applying the aforementioned conditions in Eqs.\eqref{Arenweyl} and \eqref{Arenconformalgen}, the renormalized area can be rewritten as

\begin{equation}
\mathcal{A}_{\text{ren}}\left [\Sigma \right ]= \frac{\ell ^{2}}{2}\int\limits _{\Sigma }d^{2}y\sqrt{\gamma }\left(W_{\left(\text{E}\right)ij}^{ij} -P_{ij}^{\left (A\right )}P_{\left (A\right )}^{ij}\right) -2\pi \ell ^{2}\chi \left [\Sigma \right] \,.
\label{ArenWillmore}
\end{equation}
N.B. that Eq.\eqref{ArenWillmore} corresponds to the evaluation of a codimension-2 conformal invariant, $L_{\Sigma}$, given by
\begin{equation}
L_{\Sigma}= \frac{\ell ^{2}}{2}\int\limits _{\Sigma }d^{2}y\sqrt{\gamma }\left(W_{ij}^{ij} -P_{ij}^{\left (A\right )}P_{\left (A\right )}^{ij}\right) -2\pi \ell ^{2}\chi \left [\Sigma \right] \,,
\label{Lsigma}
\end{equation}
for minimal surfaces on Einstein-AdS spacetimes which are anchored at the boundary.

\subsection{Willmore energy for Einstein ambient spaces}

A characteristic example of a functional with manifest conformal symmetry is Willmore energy. This is defined in a compact and orientable two-dimensional surface immersed in $\mathbb{R}^{3}$ \cite{Toda2017Willmore,willmore1996riemannian,marques2014willmore}. In particular, the Riemannian manifold $\mathbb{R}^{3}$ arises at the conformal frame of a constant time slice $t=const$ in a pure $AdS_4$ metric $g_{\mu \nu}$ \cite{Fonda:2015nma}. In this case the Weyl tensor vanishes identically and the $\left(A\right)$ label can be dropped, since $\mathcal{K}^{\left(t\right)}=0$. As a consequence, Eq.\eqref{ArenWillmore} depends explicitly on the square of the traceless extrinsic curvature and the Euler characteristic. Nevertheless, taking into account that in general

\begin{equation}
P_{ij}^{\left (A\right )}P_{\left (A\right )}^{ij} =\mathcal{K}_{ij}^{\left (A\right )}\mathcal{K}_{\left (A\right )}^{ij} -\frac{1}{2}\left (\mathcal{K}^{\left (A\right )}\right )^{2} \,, \label{psquared}
\end{equation}
and considering the Gauss-Codazzi relations

\begin{align}
R_{ij}^{ij} &=R +R_{ABAB}-2R_{AA} \,, \nonumber \\
& =\mathcal{R} +\mathcal{K}_{ij}^{\left (A\right )}\mathcal{K}_{\left (A\right )}^{ij} -\left (\mathcal{K}^{\left (A\right )}\right )^{2} \,,
\label{GaussCodazzi}
\end{align}
we arrive in the following expression

\begin{equation}
\mathcal{A}_{\text{ren}}\left [\Sigma \right ] = -\frac{\ell ^{2}}{2}\int \limits_{\Sigma }d^{2}y\sqrt{\gamma}\left (R^{ij}_{ij}-\mathcal{R} +2 H^{2}\right)-2\pi \ell ^{2}\chi \left [\Sigma \right] \,.
\label{Arenwillmore2}
\end{equation}
The evaluation of the latter in $\mathbb{R}^{3}$ requires moving to the conformal frame $\hat{g}_{\mu \nu}$, where the Riemann curvature vanishes identically, leading to

\begin{equation}
\mathcal{A}_{\text{ren}}\left [\Sigma \right ] = \frac{\ell ^{2}}{2}\int \limits_{\Sigma }d^{2}y\sqrt{\hat{\gamma}}\left (\hat{\mathcal{R}} -2 \hat{H}^{2}\right) -2\pi \ell ^{2}\chi \left [\Sigma \right ] \,.
\label{Arenwillmore3}
\end{equation}
Note, that up to this point we have not constrained the localization of $\Sigma$. In our construction, this can be a compact submanifold deep in the bulk or a cosmic brane anchored at the conformal boundary.

For a compact codimension-2 surface $\Sigma$, the Euler theorem

\begin{equation}
\int \limits_{\Sigma_{\text{comp}} }d^{2}y\sqrt{\hat{\gamma}}\hat{\mathcal{R}} =4\pi \chi \left [\Sigma_{\text{comp}} \right ] \,,
\label{2DEulertheorem}
\end{equation}
simplifies Eq.\eqref{Arenwillmore3} significantly, which now reads

\begin{equation}
\mathcal{A}_{\text{ren}}\left [\Sigma_{\text{comp}} \right ]= -\ell^{2}\mathcal{W}\left [\Sigma_{\text{comp}} \right ] \,, \label{conformalWillmorecompact}
\end{equation}
where

\begin{equation}
\mathcal{W}\left [\Sigma \right ] =\int\limits_{\Sigma }d^{2}y\sqrt{\gamma} H^2 \,,
\end{equation}
is the Willmore energy functional.

In the case of a boundary-anchored non-compact $\Sigma$, one has to follow the prescription proposed in Refs.\cite{Alexakis:2010zz,Fonda:2015nma}. There, a closed surface $2\Sigma$ is constructed by doubling $\Sigma$, such that $2\Sigma = \Sigma \cup \tilde{\Sigma}$, with $\tilde{\Sigma}$ being a minimal surface embedded into the mirror manifold $\tilde{M}$ beyond the conformal boundary. Taking into account the Euler theorem \eqref{2DEulertheorem} for the compact surface $2\Sigma$  and considering that the Euler characteristic behaves as $\chi \left(2\Sigma\right)=2\chi \left(\Sigma\right)$, leads to

\begin{equation}
\mathcal{A}_{\text{ren}}\left [\Sigma_{\text{non-comp}} \right ]= -\frac{\ell^{2}}{2}\int \limits_{2\Sigma }d^{2}y\sqrt{\hat{\gamma}}\hat{H}^2 \,, \label{conformalWillmore}
\end{equation}
or in terms of the Willmore energy,

\begin{equation}
\mathcal{A}_{\text{ren}}\left [\Sigma_{\text{non-comp}} \right] = -\frac{\ell^{2}}{2}\mathcal{W}\left [2\Sigma \right ] \,.
\label{ArenWillmorenoncomp}
\end{equation}
Here, Eqs.\eqref{conformalWillmorecompact} and \eqref{ArenWillmorenoncomp} make manifest the conformal invariance of Willmore energy for conformally flat ambient metrics since both arise as special cases of \eqref{ArenWillmore}.

The previous analysis made explicit the fact that, in general, the renormalized area functional is not a local conformal invariant of the codimension-2 hypersurface. However, one may restore conformal invariance in particular cases, such as for minimal hypersurfaces embedded in an Einstein spacetime.

\section{Obtaining $L_{\Sigma}$: the easy way.} \label{CGconical}

It is well-known that Einstein spacetimes are Bach-flat spacetimes. In particular, Einstein spacetimes arise as solutions of CG. Maldacena in Ref.\cite{Maldacena:2011mk} shows the emergence of Einstein-AdS gravity from 4D CG at tree level when Neumann boundary conditions for the metric are considered. The equivalence between the action of Conformal Gravity evaluated on Einstein spaces and renormalized Einstein-AdS gravity was made explicit in Ref.\cite{Anastasiou:2016jix}. The resulting Einstein-AdS action is indeed free from IR divergences. Thus, in the Einstein sector, both theories describe the same physics, what can be extended to the corresponding boundary field theories when considering the gauge/gravity duality.

\subsection{Embedding Einstein-AdS gravity in CG}



Going deeper into the rabbit-hole, one realizes the appearance of the MacDowell-Mansouri action \cite{MacDowell:1977jt} out of the renormalized Einstein-AdS action given in Eq.\eqref{IrenGB}, i.e., for that particular Gauss-Bonnet coupling. Indeed, this action  takes the form

\begin{equation}
I_{\text{ren}} = \frac{\ell^2}{256 \pi G_{N}} \int\limits_{M} d^{4}x \sqrt{g} \delta^{\mu_{1} \mu_{2} \mu_{3} \mu_{4}}_{\nu_{1} \nu_{2} \nu_{3} \nu_{4}} \left(R^{\nu_{1} \nu_{2}}_{\mu_{1} \mu_{2}} + \frac{1}{\ell^2} \delta^{\nu_{1} \nu_{2} }_{\mu_{1} \mu_{2}}\right) \left(R^{\nu_{3} \nu_{4}}_{\mu_{3} \mu_{4}} + \frac{1}{\ell^2} \delta^{\nu_{3} \nu_{4} }_{\mu_{3} \mu_{4}}\right) - \frac{\pi \ell^2}{2 G_{N}} \chi \left[M\right] \,,
\label{MacDowellMansouri}
\end{equation}
which suggests  a connection to CG, since it can be expressed as the square of the Weyl tensor for Einstein spaces, $W_{\left (E\right )\mu \nu }^{\alpha \beta }$, that is

\begin{equation}
I_{\text{ren}}
=\frac{\ell ^{2}}{64\pi G_{N}}\int\limits _{M}d^{4}x\sqrt{g}\,W_{\left (E\right )\mu \nu }^{\kappa \lambda }W_{\left (E\right )\kappa \lambda }^{\mu \nu } -\frac{\pi \ell ^{2}}{2G_{N}}\chi \left [M\right ] \,. \label{renEH}
\end{equation}
It is, therefore, an interesting possibility to consider the renormalized Einstein-AdS action as coming from CG, i.e.,

\begin{equation}
I_{\text{CG}} =\frac{\ell ^{2}}{64\pi G_{N}}\int \limits_{M}d^{4}x\sqrt{g}\,W_{\mu \nu }^{\kappa \lambda }W_{\kappa \lambda}^{\mu \nu } -\frac{\pi \ell ^{2}}{2G_{N}}\chi \left [M\right] \,. \label{IConformalGrav}
\end{equation}
This functional describes the dynamics of a higher-derivative gravity theory, as the corresponding field equations are of fourth order in derivatives. It has been shown that this action is finite for generic asymptotically AdS boundary conditions \cite{Grumiller:2013mxa}. It is then reasonable to think that, by a proper embedding of Einstein-AdS gravity in CG, the Einstein sector of the theory would inherit the cancellation of IR divergences in the radial, holographic coordinate.

\subsection{Conical contributions of CG}

In order to determine the conical contribution of CG when evaluated on the orbifold $M^{\left (\alpha\right )}$, we consider its expression in terms of the Riemann curvature given by

\begin{equation}
I_{\text{CG}}=\frac{\ell ^{2}}{64\pi G_{N}}\int \limits_{M}d^{4}x\sqrt{g}\left (Rie^{2} -2 Ric^{2} +\frac{1}{3}R^{2}\right ) -\frac{\pi \ell ^{2}}{2G_{N}}\chi \left [M\right ] \,.
\end{equation}
Then, we use the FPS expressions for the Euler characteristic and the quadratic terms in the curvature given in Eqs.\eqref{Riemsquared}-\eqref{RicciScalaquared}, obtaining

\begin{align}
I^{\left(\alpha\right)}_{CG} &=\frac{\ell ^{2}}{64\pi G_{N}}\int \limits_{M^{\left (\alpha \right )}}d^{4}x\sqrt{g}\left \vert W^{\left (\alpha \right )}\right \vert ^{2} -\frac{\pi \ell ^{2}}{2G_{N}}\chi \left [M^{\left (\alpha \right )}\right ] \,,
\end{align}
where

\begin{equation}
\int\limits_{M^{\left(\mathcal{\alpha}\right)}} d^{4}x \sqrt{g}\left \vert W^{\left (\alpha \right )}\right \vert ^{2} = \int\limits_{M} d^{4}x \sqrt{g} \left \vert W\right \vert ^{2} +8\pi \left (1 -\alpha \right ) \int\limits_{\Sigma} d^{2}y \sqrt{\gamma}K_{\Sigma } +\mathcal{O}\left (\left (1 -\alpha \right )^{2}\right ) \, \label{ICGconicalexpa}
\end{equation}
denotes the expansion of the Weyl squared term in the conical parameter.
In the previous expression,
\begin{equation}
K_{\Sigma } =R_{ABAB} -R_{AA} +\frac{1}{3}R +\frac{1}{2}\left (\mathcal{K}^{\left (A\right )}\right )^{2} -\mathcal{K}_{ij}^{\left (A\right )}\mathcal{K}_{\left (A\right )}^{ij} \label{Ksigma1}\,
\end{equation}
is a conformal covariant term on the 2D manifold $\Sigma $ endowed with the metric $\gamma _{ij}$ \cite{Solodukhin:2008dh}. Indeed, one can show that $K_{\Sigma }$ consists of the sum of two objects: i) the subtraces on $\Sigma $ of the bulk Weyl tensor and ii) the square of the traceless part of the extrinsic curvature. In particular, the intrinsic curvature terms can be resumed as

\begin{equation}
W_{ij}^{ij} =R_{ABAB} -R_{AA} +\frac{1}{3}R \,, \label{wabab}
\end{equation}
whereas the extrinsic curvature terms are given in Eq.\eqref{psquared}
such that Eq.\eqref{Ksigma1} can equivalently be cast in the form

\begin{equation}
K_{\Sigma }=W_{ij}^{ij} -P_{ij}^{\left (A\right )}P_{\left (A\right )}^{ij} \,.
\label{KSigma}
\end{equation}
As it has been shown in the previous section, this is a conformally covariant combination of weight -2. In this case however, no restrictions have been imposed on the class of surfaces or spacetimes for which the last formula is valid for. As a consequence, the conical parameter expansion of the CG action in the orbifold becomes

\begin{equation}
I_{\text{CG}}^{\left(\mathcal{\alpha}\right)} = I_{\text{CG}} + \frac{\left(1-\alpha\right)}{4 G_{N}} L_{\Sigma} +\mathcal{O}\left (\left (1 -\alpha \right )^{2}\right )
\end{equation}
where the first order conical contribution obtains the form

\begin{equation}
L_{\Sigma } = \frac{\ell^{2}}{2}\int \limits_{\Sigma }d^{2}y\sqrt{\gamma}K_{\Sigma } - 2\pi \ell^{2} \chi \left [\Sigma \right]   \label{ICGconical} \,.
\end{equation}
Hence, it becomes clear that the manifest conformal symmetry of the bulk action is induced on the cod-2 functional constructed by the conical contributions. Note that neither the shape nor the compactness of $\Sigma$ are constrained. As a consequence, $L_{\Sigma}$ refers to an arbitrary two-dimensional surface immersed in a generic four-dimensional metric. When AdS asymptotics are considered and $\Sigma$ is a surface anchored at the boundary, then $L_{\Sigma}$ is interpreted as the HEE for CFTs dual to Conformal Gravity.

Interestingly enough, the last expression \eqref{ICGconical} reduces to the conformally invariant form of the renormalized area \eqref{ArenWillmore} when evaluated on Einstein spacetimes. However, the former is valid for a generic surface $\Sigma$ whereas the latter applies only to minimal surfaces. The relation of the $L_{\Sigma}$ functional to the renormalized area along with its manifest conformal invariance is a feature of having considered conformal invariance in the bulk as the starting point. This codimension-2 conformal invariant will be further applied to derive other energy functionals in Section \ref{ReducedHawking}.

\subsection{Willmore energy}

The conformal invariance of the functional $L_{\Sigma}$ suggests that one should be able to recover Willmore energy at the proper limit. Our starting point in this analysis, is the Weyl contribution of Eq.\eqref{ICGconical}, which is decomposed as

\begin{equation}
W_{ij}^{ij} =R_{ij}^{ij} -2S_{i}^{i} \,.
\end{equation}
This form comes from the codimension-2 sub-trace of Eq.\eqref{Weyltensor}. Furthermore, taking into account the Gauss-Codazzi relation of Eq.\eqref{GaussCodazzi} along with Eq.\eqref{psquared}, the $L_{\Sigma}$ functional can be rewritten as

\begin{equation}
L_{\Sigma } =\frac{\ell^{2}}{2}\int \limits_{\Sigma }d^{2}y\sqrt{\gamma }\left [\mathcal{R} -2\left (H^{\left (A\right )}\right )^{2} -2 S_{i}^{i}\right ] -2\pi \ell^{2} \chi \left [ \Sigma \right ] \,.
\end{equation}
Making contact with Willmore energy requires the hypersurface $\Sigma$ to be compact, in order to simplify the integral of the intrinsic Ricci scalar (Gauss-Bonnet density) and the Euler characteristic using the Euler theorem of Eq.\eqref{2DEulertheorem}. This consideration yields

\begin{equation}
L_{\Sigma_{\text{comp}}} = - \ell^{2}\int \limits_{\Sigma }d^{2}y\sqrt{\gamma}\left [\left (H^{\left (A\right )}\right )^{2} +S_{i}^{i}\right ] \,. \label{ICG3}
\end{equation}
When the surface $\Sigma$ is embedded an arbitrary Riemannian manifold $\mathcal{M}_3$, defined in a constant time slice of the AAdS bulk, the above expression is interpreted as the Conformal Willmore energy \cite{Mondino_2018}. In the case of a pure AdS bulk, the latter reduces to the standard Willmore energy functional when embedded in $\mathbb{R}^3$. As has been discussed in the previous section, this is achieved by considering the unphysical conformal frame $\hat{g}$ where the metric is that of $\mathbb{R}^3$ and therefore the transformed Schouten tensor vanishes identically, leading to

\begin{align}
L_{\Sigma_{\text{comp}}}\left[\mathbb{R}^3\right] & = -\ell^{2}\int \limits_{\Sigma_{\text{comp}} }d^{2}y\sqrt{\hat{\gamma}} \hat{H}^{2} \nonumber \\
&=-\ell^{2} \mathcal{W}\left [\Sigma \right] \,.
\label{Lsigmacompact}
\end{align}
When non-compact surfaces anchored at the conformal boundary are considered then one can use the doubling construction (discussed in subsection \ref{Willmoreenergy}). In this case, it is straightforward to show that

\begin{equation}
L_{\Sigma_{\text{non-comp}}} \left[\mathbb{R}^3\right]=-\frac{\ell^2}{2} \mathcal{W}\left [2\Sigma \right] \,.
\label{Lsigmanoncompact}
\end{equation}
Note that the renormalized area $\mathcal{A}_{\text{ren}}$ and the Willmore energy $\mathcal{W}$ functionals can be obtained as particular cases of the conformal invariant $L_{\Sigma}$. Point in fact, $\mathcal{A}_{\text{ren}}$ is recovered when considering Einstein spacetimes and minimal boundary-anchored surfaces, and $\mathcal{W}$ is obtained for compact --or doubled-- surfaces in pure AdS. This is reminiscent of the equivalence in the bulk, between CG and Einstein-AdS gravity when the former is evaluated in Einstein spacetimes.

The conical contribution in Eq.\eqref{ICGconicalexpa} is general and it will be used in the following section to derive the reduced Hawking mass from CG.

\section{Reduced Hawking mass from $L_{\Sigma}$}\label{ReducedHawking}

The results of the previous section show how starting from the codimension-2 conformal invariant $L_{\Sigma}$, one can derive the renormalized area and the Willmore energy functionals. This is achieved by imposing certain restrictions on both the ambient spacetime and the codimension-2 surface under consideration. 

Nonetheless, $L_{\Sigma}$ is defined for generic surfaces. Thus, it is expected that it could provide a generalization of renormalized area for non-extremal surfaces in the Einstein limit, such as surfaces that are not orthogonally anchored at the AdS boundary.

Our starting point is the conical contribution of the CG action, given in Eq.\eqref{ICGconical}. Evaluating this expression for Einstein spacetimes amounts to the replacement of the Weyl tensor with $W_{\left (E\right )\mu \nu }^{\alpha \beta }$. Thus, one gets that

\begin{equation}
L_{\Sigma }\left [E\right ] = \frac{\ell^{2}}{2}\int \limits_{\Sigma }d^{2}y\sqrt{\gamma}\left (W_{\left (E\right )ij}^{ij} -P_{ij}^{\left (A\right )}P_{\left (A\right )}^{ij}\right ) -2\pi \ell^{2} \chi \left [\Sigma \right] \,,
\label{LEinstein}
\end{equation}
which, taking into account Eqs.\eqref{psquared} and\eqref{GaussCodazzi}, can be cast in the form

\begin{equation}
L_{\Sigma }\left [E\right ] = \frac{\ell^{2}}{4}I_{H}\left [\Sigma \right ]-2\pi \ell^{2} \chi \left [\Sigma \right] \,, \label{ICGconical2} 
\end{equation}
where

\begin{equation}
I_{H}\left [\Sigma \right ] =2\int \limits_{\Sigma }d^{2}y\sqrt{\gamma}\left [\mathcal{R} +\frac{2}{\ell ^{2}} -\frac{1}{2}\left (\mathcal{K}^{\left (A\right )}\right )^{2}\right ] \,, \label{hawkingmass}
\end{equation}
is the generalization of the Hawking mass for AAdS spaces, which is referred to as reduced Hawking mass in Ref.\cite{Fischetti:2016fbh}. The authors in that reference exhibit the finiteness of this object when an arbitrary boundary-anchored hypersurface $\Sigma $ in a constant time slice is considered.

Further properties of the reduced Hawking mass can be worked out by rewriting Eq.\eqref{ICGconical2} as

\begin{equation}
L_{\Sigma }\left [E\right ] =\mathcal{A}_{\text{ren}}\left [\Sigma \right ] -\frac{\ell ^{2}}{4}\int \limits_{\Sigma }d^{2}y\sqrt{\gamma}\left (\mathcal{K}^{\left (A\right )}\right )^{2} \,, \label{ICGconical3}
\end{equation}
in terms of the renormalized area $\mathcal{A}_{\text{ren}}\left [\Sigma \right ]$ of an arbitrary two-dimensional hypersurface $\Sigma $, anchored orthogonally to the boundary \cite{Alexakis:2010zz}.  It is then manifest that $L_{\Sigma }\left [E\right ]$ becomes proportional to the renormalized area $\mathcal{A}_{\text{ren}}\left [\Sigma \right ]$, when $\Sigma $ is minimal.

Eq.\eqref{ICGconical3} indicates that $\mathcal{A}_{\text{ren}}$ diverges when $\Sigma $ is anchored to the boundary at an arbitrary angle. It is the functional $I_{H}\left [\Sigma \right ]$ the one that correctly cancels the divergences in the most general case.
As a consequence, the reduced Hawking mass generalizes the concept of renormalized area to non-minimal hypersurfaces.

The different energy functionals obtained from $L_{\Sigma}$, with the corresponding restrictions on the ambient space $M$ and on the codimension-2 surface $\Sigma$, are shown in Table \ref{Table 1}.

\begin{table}[H]
    \centering
\begin{tabular}{|l|c|c|}\hline
\diagbox[width=10em]{$\Sigma$}{$M$}&
  Einstein &  pure AdS \\ \hline
 min & $\mathcal{A}_{\text{ren}}$ & $\mathcal{W}$\\ \hline
    non-min & $I_{H}$ &  \\ \hline
\end{tabular}
\caption{Energy functionals from $L_{\Sigma}$}\label{Table 1}
\end{table}

\section{Conclusions}\label{Conclusions}
In this work, we have obtained a local conformally invariant object in codimension-2 $L_{\Sigma}$, from four-dimensional CG evaluated on the replica orbifold using the FPS relations \cite{Fursaev:2013fta}, which inherits its conformal symmetry from the bulk. $L_{\Sigma}$ reduces to energy functionals which generalize renormalized area, such as reduced Hawking mass and Willmore energy, for generic Einstein or pure AdS ambient spaces respectively. Said functionals may be used for codimension-2 surfaces which are boundary-anchored at an arbitrary angle, whether they are minimal or not.

\begin{figure}[h]
\includegraphics[width=\textwidth]{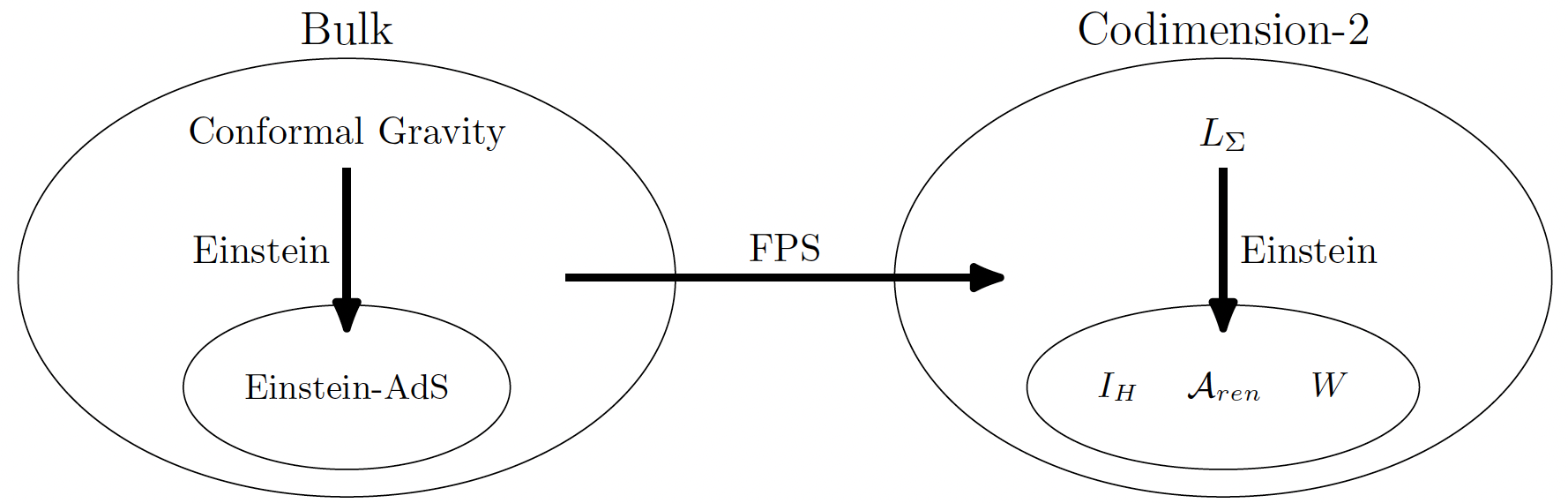}
\caption{Diagram showing the relations between the functionals discussed in the paper.}
\label{Figure 1}
\end{figure}

The relations between the functionals, both at the bulk and at the codimension-2 levels, as well as the fact that they are embedded in conformal invariant structures, are summarized in Figure \ref{Figure 1}.

The presented procedure streamlines the derivation discussed in Ref.\cite{Anastasiou:2020smm}, where the conformal invariant was obtained starting from the renormalized area of minimal surfaces given in Ref.\citep{Alexakis:2010zz}.
Then, conformal invariance was restored by rewriting the expression in a manifestly invariant form, for restricted Einstein spacetimes, as shown in Section \ref{Section 2}.

We establish the fundamental role of the bulk conformal symmetry in the renormalization of geometrical structures residing in codimension-2 submanifolds. The new result is the construction of $L_{\Sigma}$ out of the conical contributions of CG. In this way, this functional has a manifest local conformal symmetry, acquired from the bulk action. Most importantly, when AAdS Einstein spacetimes are considered and for a surface $\Sigma$ anchored at the boundary, we recover the reduced Hawking mass. This quantity is finite for an arbitrary boundary anchored 2D hypersurface, and it is monotonous under an inverse mean curvature flow, as shown in Ref.\cite{Fischetti:2016fbh}. This fact highlights the role of conformal symmetry in the renormalization procedure. As the functional can be obtained from $L_{\Sigma}$, a possible reinterpretation of the monotonicity property is suggested.

The procedure here presented opens the possibility of the derivation of generalized energy functionals in higher dimensions, starting from conformal invariance in the bulk.

\section{Acknowledgments}
We thank Marika Taylor and Nicolas Boulanger for interesting discussions. We also thank Prof. Boulanger for his kind hospitality at U. Mons during the completion of this work. We appreciate the help of Andrés Argandoña in designing Figure 1. The work of GA is funded by  ANID, Convocatoria Nacional Subvenci\'on a Instalaci\'on en la Academia Convocatoria A\~no 2021, Folio SA77210007. The work of IJA is funded by ANID, REC Convocatoria Nacional Subvenci\'on a Instalaci\'on en la Academia Convocatoria A\~no 2020, Folio PAI77200097. The work of RO is funded by ANID Grant N$^{\circ }$1190533 {\it Black holes and asymptotic symmetries}, and ANID/ACT210100 Anillo Grant  {\it Holography and its applications to High Energy Physics, Quantum Gravity and Condensed Matter Systems}.

\appendix

\section{Notation} \label{AppendixA}

In Table \ref{tab:notation}, we present the notation used for the coordinates, metric and curvature terms --intrinsic and extrinsic-- for the different manifolds considered in the paper.

\begin{table}[H]
    \centering
    \begin{spacing}{2}
    \begin{tabular}{|l|c|c|c|}
     \cline{2-4}
    \multicolumn{1}{c|}{}
 & $\mathcal{M}$ &  $\partial\mathcal{M}$ & $\Sigma$\\ 
 \hline\hline
        Indices & $\mu,\nu$ &$a,b$& $i,j$ \\
    Coordinates & $x^\mu$ &$X^a$& $y^{i}$ \\
    Metric & $g_{\mu\nu}$ & $h_{ab}$ & $\gamma_{ij}$\\
    Riemann tensor & $R_{\mu\nu}^{\lambda\sigma}$ & $\frak{R}_{ab}^{cd}$ & $\mathcal{R}_{km}^{ij}$ \\
    Extrinsic curvature & & $K_{ab}$ &$\mathcal{K}_{ij}$\\ \hline
    \end{tabular}
    \end{spacing}
    \caption{Notation for the geometric objects indicated in the first column, for the different manifolds considered in the first row.}\label{tab:notation}
\end{table}

\bibliographystyle{JHEP-2}
\bibliography{bibliography}

\end{document}